# Linguistic Dead-Ends and Alphabet Soup: Finding Dark Patterns in Japanese Apps

**SHUN HIDAKA**

*Tokyo Institute of Technology*
*Tokyo, Japan*

**SOTA KOBUKI**

*Tokyo Institute of Technology*
*Tokyo, Japan*

**MIZUKI WATANABE**

*Tokyo Institute of Technology*
*Tokyo, Japan*

**KATIE SEABORN**

*Tokyo Institute of Technology*
*Tokyo, Japan*

**ABSTRACT:**  Dark patterns are deceptive and malicious properties of user interfaces that lead the end-user to do something different from intended or expected. While now a key topic in critical computing, most work has been conducted in Western contexts. Japan, with its booming app market, is a relatively uncharted context that offers culturally- and linguistically-sensitive differences in design standards, contexts of use, values, and language, all of which could influence the presence and expression of dark patterns. In this work, we analyzed 200 popular mobile apps in the Japanese market. We found that most apps had dark patterns, with an average of 3.9 per app. We also identified a new class of dark pattern: "Linguistic Dead-Ends" in the forms of "Untranslation" and "Alphabet Soup." We outline the implications for design and research practice, especially for future cross-cultural research on dark patterns.

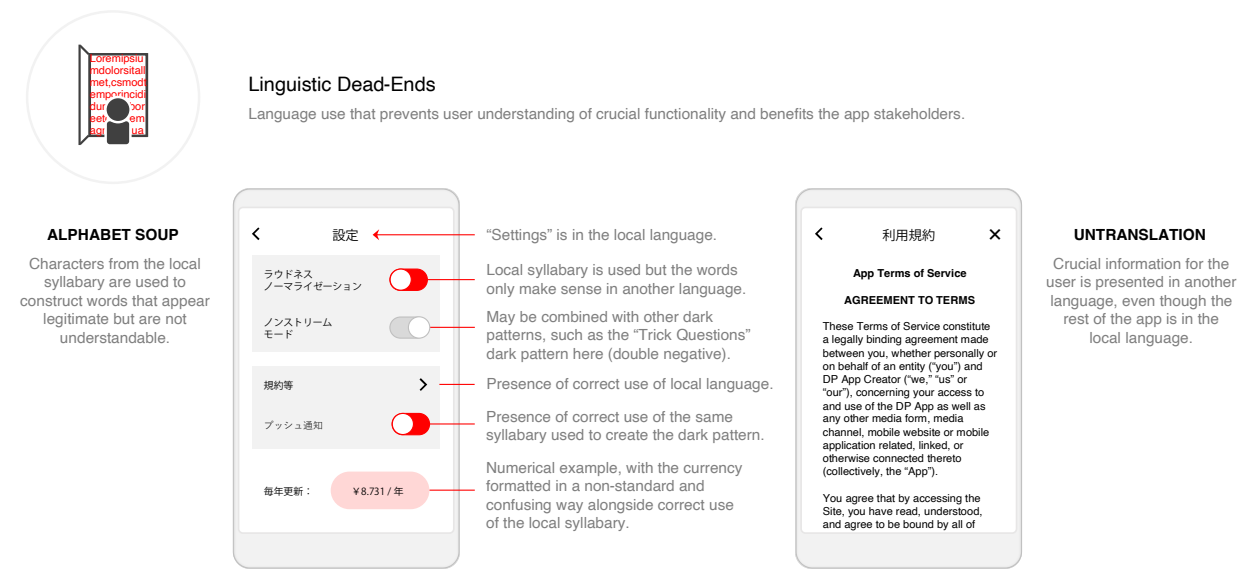

**Figure 1:** *A new class of dark pattern, Linguistic Dead-Ends, and its subclasses, Alphabet Soup and Untranslation, identified through a cultural- and language-sensitive approach within the Japanese app market context.*

## 1 Introduction

Dark patterns (DPs) are a type of malicious interface design pattern that tricks or forces the end-user into taking an action that is different from what they intend or expect, one that benefits the purveyor of the interface. DPs have rapidly gained attention in human-computer interaction (HCI) research and practice [10, 16, 32]. Dark patterns are a topic that should concern everyone involved with websites and apps: the users who may be affected by them, the researchers who should sound the whistle on them, the designers who unintentionally create them, and everyone involved with websites and apps. In recent years, research has been conducted that directly involves designers [15], under government supervision [40], and from a combined perspective of law, ethics, computer science, HCI, and others [17]. Previous research on DPs includes the proposal of a DP taxonomy [16, 5, 8], the analysis of DPs in different categories of apps [9], the analysis of DPs in specific contexts and the creation of a taxonomy from these designs [28], and research related to user perception of DPs [9]. Several studies on the presence, frequency, and user impact of dark patterns have been conducted, mainly in Europe and the United States, i.e., Western contexts, where DPs have been grouped together to create distinct and easily understood classes[1], and axes have been established to determine if they are DPs to be used in various studies, such as DP distribution studies and user research. However, Eastern contexts, such as Japan, may approach DPs differently or not at all. Cultural differences, including with respect to design in general and specifically within e-commerce contexts, as well as language differences, may influence whether and how DPs are employed. As such, the presence, frequency, classes, and features of DPs within the Japanese context may differ from what has been discovered in the Western research so far. At present, there appears to have been no research on DPs in Japan.

Opportunities for the Japanese public to encounter UIs that employ DPs are increasing every year. For example, according to research by Japan's Ministry of Internal Affairs and Communications, the amount of time spent online per day on weekdays for all age groups in Japan increased by 39.6% over five years from 2015 to 2019[2]. Furthermore, research by Japan's Ministry of Economy, Trade and Industry indicates that the size of the e-commerce market has expanded by about 40.6% over five years from 2015 to 2019[3]. While no research to date has evaluated the Japanese context, the booming e-commerce market and widescale use of mobile devices and apps raises grave concerns about the existence and extent to which DPs influence Japanese consumers. To limit these risks and prevent harm, it is important to understand what DPs, if any, are being employed, to what extent, and how they might impact end-users. Moreover, if there are DPs particular to the Japanese cultural context and language, they are as yet unknown. However, marking out scale and effects of DPs within the Japanese context will require addressing these potential, hidden DPs.

To this end, we conducted the first study on DPs in Japan. We aimed to clarify the distribution and frequency of dark patterns in Japan for researchers, designers, and users. We also aimed to analyze and clarify dark patterns from three perspectives: exploring the Japanese context with comparison to previous studies; determining any differences in dark patterns over time; and analyzing results based on differences in the devices used. As such, we sought to answer the following three research questions: *(RQ1) What kinds of dark patterns are present in Japanese apps? (RQ2) How are they distributed by class, subclass, and across app categories? (RQ3) What differences are there in the presence and/or frequency of dark patterns between the Japanese*

---

[1] For consistency and to avoid confusion between DPs and where the apps were sourced from ("app categories"), we refer to high-level DP categories as "classes" (rather than "types" or "categories") and DPs under these classes as "subclasses" (rather than "subtypes" or "subcategories").
[2] https://www.soumu.go.jp/johotsusintokei/whitepaper/ja/r02/html/nd252510.html
[3] https://www.meti.go.jp/press/2021/07/20210730010/20210730010.html

*and the US app markets, if any?* To answer these questions, we replicated an existing dark patterns research protocol for the context of Japan. Specifically, we used the qualitative expert classification protocol of Di Geronimo et al. [9]. For this, we analyzed a total of 200 apps (25 apps in each of eight app categories) and classified the DPs we found into six classes. Five classes were devised by the formative research of Gray et al. [16] and Di Geronimo et al. [9], and one class, Linguistic Dead-Ends, was newly discovered and defined by us during the course of this study. Our work has three main contributions:

(1) Discovery of a novel class of DP not previously found, which we have termed the "Linguistic Dead-End," and its two subclasses, "Untranslation" and "Alphabet Soup," that may be found outside of the Japanese context;
(2) Empirical evidence of DPs across app categories that align with the DP taxonomy devised by Gray et al. [16] and refined by Di Geronimo et al. [9], thereby extending an established model of DPs to the context of Japan; and
(3) A cross-cultural comparison of findings from previous research in Western and/or English-speaking app contexts, indicating significantly less representation of DPs in the Japanese context, specifically an average of 3.9 compared to 7.4.

This work acts as a bridge in critical computing and interaction design between the East and the West. It is expected to influence how research on DPs is conducted in Japan and around the world, especially in terms of being sensitive to differing cultural and language contexts.

## 2  Background

Critical computing scholarship has pointed to ethical issues in the design of modern interfaces and technologies [6]. Dark patterns have subsequently become a feature of the discourse within HCI spaces, both academic and otherwise [10, 32]. We review the work so far and discuss how we extend it to the case of the Japanese app market. We also justify our positioning within a larger shift towards inclusive and intersectional research in HCI [22, 24, 35].

### 2.1   DARK PATTERNS AND MALICIOUS INTERFACES

Dark patterns have become a hallmark of research on ethical design practices and ethical HCI. The term "dark patterns" was coined by Brignull in 2010 when he registered darkpatterns.org [8] (now called Deceptive Design). Dark patterns are part of the user interface (UI), typically as visual interactive elements embedded within a user function or responsive to user engagement. For instance, the Preselection DP [16] is a design pattern in which checkboxes for app notification settings and permission to receive newsletters are checked before they are selected by the user, which can lead to the delivery of unsolicited notifications and newsletters that the user does not want. Since 2010, ethical designers and researchers in HCI and interaction design have focused on finding new classes of DPs [5, 11, 16], studying the distribution of DPs across websites and apps in specific contexts [3, 9, 28], investigating user perceptions and reactions to DPs [4, 9, 11, 14, 26, 30, 40], and better understanding the perspective of designers on DPs [15]. Building on Brignull's initial examples of DPs, Bösch et al. [5] created classes related to privacy. Gray et al. [16] then devised a hierarchical taxonomy consisting of five classes: Nagging, Obstruction, Sneaking, Interface Interference, and Forced Action. These were then refined by Di Geronimo et al. [9]. The variety and nature of DPs are expected to evolve as research progresses and technology changes—and new contexts are explored. To this end, we use the version of the DP taxonomy created by Gray et al. [16] and refined by Di Geronimo et al. [9] as a starting point to explore the uncharted case of Japan.

Much of the work so far has focused on determining the existence and distribution of DPs across a range of UIs and contexts of use. In their study of the distribution of DPs, Mathur et al. [28] used an automated crawling system to explore DPs on 11K shopping sites and found that DPs were present on at least 11.1% of the sites, and that the higher the site's Alexa ranking, the more likely a DP was to appear. DPs have also been explored in browser cookies, which temporarily store user information on websites. Krisam et al. [21] found that many of the 500 most popular websites in Germany used a deceptive design that visually led the user to "allow all" cookies, attempting to achieve the designer's intent in a natural way, even in UIs that involve user privacy. Di Geronimo et al. [9] detected DPs in the recordings of user interactions across 240 free apps in eight major categories downloaded from the US-based Google Play store. They focused on the 30 top-ranking apps per category, using a predefined procedure to find the types and number of DPs that appeared in the app UIs. They showed that as many as 95% of the apps used at least one DP, with an average of 7.4 DPs per app category. We built on this work by considering the app market in Japan, which has not yet been explored.

More and more research has been conducted on people's reactions to DPs. Bongard-Blanchy et al. [4] found that generation z, y, and x, in that order, were more likely to recognize DPs; moreover, college graduates were more likely to recognize DPs than those who had not graduated college, i.e., age and education were correlated with the ability to identify certain DPs. In Di Geronimo et al. [9], people of different nationalities and backgrounds from over 40 countries were recruited to identify DPs. Of these, 25% noticed malicious designs and 24% were able to correctly detect DPs. This suggests that most people will not be able to recognize DPs and other designs of malicious intent correctly. Gray et al. [14], through a cross-cultural and cross-geographic survey of "manipulation" experiences in English- and Chinese-speaking countries, found that more than 80% of respondents had experienced distrust while using mobile apps and websites. Even when people feel unease, they may be unable to recognize DPs when they experience them, a phenomenon known as "DP-blindness" [9, 25]. As Di Geronimo et al. [9] discovered, experts are not immune. As such, we have taken measures to avoid DP-blindness, specifically by evaluating apps for DPs in pairs.

## 2.2 INTERSECTIONALITY AND WEIRD RESEARCH FOR ETHICAL HCI

Ethical design recognizes design as an act of power, where designers, developers, and purveyors of technology aim to influence the user experience (UX), embed values into interfaces and interactions, and should reflect on what responsibility they have to those on the receiving end [12, 18, 36]. DPs and by extension DP research are no exception. Critical explorations of DPs are a social good. However, a reflexive, critical lens raises an issue about the work conducted so far. Since the introduction of DPs by Brignull in 2010 [8] and the resulting take-up in HCI and adjacent spaces, DP research has become an active area of study in the West. Research on DPs has been conducted in or using material from the US [3, 9, 16, 25, 28, 40], Germany [5, 21], the UK [4, 11], Italy [30], and French-, Italian, and English-speaking European nations [17]. However, to the best of our knowledge, DP research elsewhere, notably in Japan and other Eastern nations with comparable industries, is virtually non-existent. One known exception is from India [3] and another may be the study by Di Geronimo et al. [9], but not enough details are provided in the paper about the 46 nationalities recruited.

We can characterize this state of affairs as a form of sampling bias called WEIRD: "Western," "educated," "industrial," "rich," and "democratic" nations [20]. While originally conceptualized by Henrich, Heine, and Norenzayan within the broader scope of behavioural research [20], WEIRD sampling biases have been found across many areas of study, including HCI and notably CHI [24]. Critical voices and initiatives have called for an expansion of *who* we include in research [22, 31, 35, 37]. While certainly a matter of inclusion and representation, it is also a matter of rigour: we cannot necessarily generalize from one population to the other, and at least we should check. Indeed, Henrich, Heine, and Norenzayan's seminal work showed that generalizing across populations is risky if not outright incorrect [19, 20]. Moreover, there are sure to be a range of populations experiencing DPs unawares and subject to any resulting harms, as yet unknown and untraceable. From an ethical computing standpoint, we must take action. Japan, as we discussed above, may be an ideal starting point, as a nation with a profile similar to those nations previously targeted in DP research. Japan is an Eastern nation with a technically savvy population and large-scale app industry that includes offerings from abroad, such as Apple, Google, and Amazon marketplaces. Exploring the case of Japan, with its differences and similarities to the nations so far included in DP research, may provide new and comparable insights, especially in the two major ways in which it differs: culture and language. We offer this work as a first step.

## 3   Methods

Our goal was to replicate the qualitative expert classification study by Di Geronimo et al. [9] for the Japanese app context so as to clarify the similarities and differences in the subclasses and distribution of DPs based on cultural and linguistic dimensions. This protocol was registered on OSF[4] before data collection and analysis on December 14$^{th}$, 2021, with an update on January 4$^{th}$, 2022 to account for a slight change in procedure regarding the third rater.

### 3.1   PROCEDURE

We aimed to replicate the procedure of the first study conducted by Di Geronimo et al. [9]. However, we made some modifications due to time, budget, and device limitations, as well as to account for special properties of the Japanese app market context, described below. The general procedure followed the original study. We started by making screen recordings of the apps, according to the procedure described in the Recording Methodology section below. Then, as per the Classification Methodology procedure, pairs of researchers identified the DPs in the recordings as a pair to account for potential DP blindness [9, 25], as well as to identify new DPs or forms of DPs related to the Japanese app context. We then analyzed the distribution of the DPs by class and app category. We describe our procedure in detail next.

---

[4] https://osf.io/jfs36

## 3.2  CORPUS GENERATION

Three researchers gathered 200 apps from the Google App Store. 24 apps were selected based on their top Google Play ranking as of September 25th, 2021, and 176 apps were selected based on their top Google Play ranking as of April 12th, 2022, using Sensor Tower as a reference. The apps were selected based on the following three criteria:

1. Must be available on the iOS platform and can be used on an iPad
2. Must be free to download
3. Must be trending in the Japanese market

We used Sensor Tower to, as much as possible, align the Japanese app categories from the Google Play rankings with those sourced in the Di Geronimo et al. [9] study. We used iPads because we were able to procure Apple devices more quickly than other devices; notably, the usage ratio of iOS and Android devices in Japan was not significantly different at the time of this research[5]. Google Play has 34 major categories of apps. Following Di Geronimo et al. [9], we tested 200 apps in eight major categories: Communication, Entertainment, Family, Music & Audio, News & Magazines, Photography, Shopping, and Social. Similarly, we also excluded certain apps based on the following six criteria:

1. Apps not currently available on the iPad
2. Apps not available in Japan
3. Apps that have already appeared in other categories
4. Apps that are launchers
5. Apps that are tied to a phone number and only one account per person is allowed
6. Apps for which the main functions of the application can only be used by logging in as a paying member

## 3.3  RECORDINGS

Recording app use by video screen capture was done by three researchers. Three iPad Air (4th generation) were used for this, each installed with either iOS 15.5 or iOS 14.8 (resulting from an iOS update during the research). Researchers either created a new Apple ID using a new email address created solely for this research or their own email address at the university. These were used to download and create accounts in the apps, as needed. The screen recording function of iOS was used to screen record the apps in use. For this, researchers started the screen recording just before app launch. All apps were recorded for 10 minutes. During these 10 minutes, the researcher performed the following six operations in this order, if they were available:

1. Create an account and log out
2. Close and reopen the application
3. Visit the product list page
4. Visit the settings page
5. Select the name of the product
6. Use the app according to its intended usage

---

[5] https://www.kantarworldpanel.com/global/smartphone-os-market-share

This cross-app operation ensured consistency in the procedure. However, it was not possible to discover dark patterns that do not appear within the first 10 minutes of typical app operation. None of the apps used were opened before the start of the screen recording. Additionally, we did not purchase any products or services through the apps. For e-shopping apps, there was a possibility that dark patterns could be found during the purchase stage, but we did not go that far and only went through the process just before completion of a purchase. Finally, for apps that required subscribing, we subscribed to the free version.

### 3.4 CLASSIFICATION PROCESS

After the Recording Method activity, pairs of three researchers watched recordings of a total of 200 apps together and discussed the DPs that they found. If the two could not agree on whether a particular design was a DP or not, and/or which DP class it fell into, we asked a fourth researcher to break the tie, as in Di Geronimo et al. [9]. This method was used for all 200 apps in this study. We aimed to make rational, consensus-driven decisions grounded in the possibility of finding DPs not found in previous studies. We did not count reappearances of the same DP that had already been identified at least once in the app. That is, if both the design of the DP and the context of its appearance were the same as a pattern that had already appeared, we recognized it as the same DP and did not count it as a new instance of a DP. This was done to ensure that the way the app was operated was not a factor in the resulting number of DPs counted within that app. When different DPs appeared in the same use case or when the same DP appeared under different operations (e.g., a pop-up mode when opening the app or a pop-up unexpectedly appearing when operating the app), each design was identified as a different DP. When the same subclass of DP appeared in succession during a sequence of operations, it was recognized as one DP if the transition was fixed as a flow, i.e., the user could not change the app operation. If it was not fixed, and there was room to open different screens depending on the user's operation in the middle of the transition screen, we counted each as a unique DP instance.

## 3.5 TAXONOMY

We used the DP classes and subclasses proposed by Gray and colleagues [16] for classification, and then fit each of the DPs we detected to the subclasses later refined by Di Geronimo and colleagues [9]. Although well-described and applicable to most of the DPs that we found, we needed to extend these classes to fit the Japanese context. Specifically, we found a new class of DP, as well as subclasses. We briefly describe the new DP that we discovered here. We discuss findings related to this new DP in the Results and Discussion sections later.

> **Nagging** refers to designs that use pop-ups or other means to persistently engage the user more than once to realize the designer's intentions.
>
> **Obstruction** refers to an UI in which some barrier makes it difficult for the user to perform the desired operation smoothly. This includes Intermediate Currency, which allows users to make transactions using only the in-app currency, not cash; Price Comparison Prevention, which makes it difficult to compare product prices with other sites by making it impossible to copy product names; and the Roach Motel, which makes it easy to get an account but difficult to delete it or exit it.
>
> **Sneaking** is a form of design in which the designer attempts to make a profit by sneaking in information relevant to the user without the user noticing. This include Bait and Switch, which causes the user to do something different from his/her expected function, Hidden Costs, which adds a portion of the price that was not shown to the user just before the purchase, Sneak into Basket, where unwanted items are mixed into the cart, and Forced Continuity, where the user is not notified when the free or trial period ends.
>
> **Interface Interference** is a design that tries to trigger a particular operation by making it easier or harder to draw attention to a particular UI element. This includes Hidden Information, which displays small or grayed-out options that are not acceptable to make users accept terms of use and other conditions; Preselection, in which undesirable options for users are selected even before they are manipulated; and Aesthetic Manipulation, in which the use of the UI disrupts users' concentration. The child classes of the Aesthetic Manipulation class include Toying with emotions, which uses UI design to induce specific actions, False Hierarchy, which makes one or more choices stand out among multiple choices, and Trick Questions, which tries to mislead the user into making a decision by using double negatives.

**Forced Action** is a design in which users who are trying to perform a particular operation are given different tasks to go through to do so. It includes Social Pyramid, in which users can earn rewards by inviting friends to the app, Privacy Zuckering, in which users are forced to provide more information than necessary, and Gamification, in which users are forced to perform tasks that are difficult to accomplish without paying.

**Linguistic Dead-End** (Figure 1) is a new DP that we discovered, referring to a design that prevents or discourages users from understanding the interface and/or content of an app by manipulating language at a fundamental level. This DP has two subclasses: **Untranslation** and **Alphabet Soup**. *Untranslation* DPs act as a language barrier by using languages other than the language of the user for part or all of the app so as to persuade or prevent understanding of essential content or functions. *Alphabet Soup* refers to information in the interface or content that is expressed using letters, numbers, and/or symbols from the local syllabary but without using real, understandable words or phrases from that syllabary, even when these exist and when the app otherwise correctly employs the local language, making the meaning difficult or impossible to grasp.

We used all six DP classes, treating the new Linguistic Dead-End DP as an addition to Gray et al.'s [16] taxonomy. We discuss the new DP and its subclasses in detail with examples in the Results and Discussion sections.

### 3.6   DEVIATIONS FROM THE ORIGINAL PROTOCOL

Our methodology differed from that Di Geronimo et al. [9] in a few ways that we can classify under two broad headings.

### 3.6.1 Changes to the Recording Methodology.

We used the lab-owned devices available only at the university, which were iPad Airs, rather than Google devices. As such, we had to use the Apple Store rather than the Google Store. Moreover, apps not compatible with iOS were excluded. Screen recording was done using the screen recording function in iOS, although we do not expect this to substantially deviate from screen recording functions in other devices. We gathered a total of 200 apps, slightly less than the 240 analyzed by Di Geronimo et al. [9], due to time and resource constraints within the research team.

### 3.6.2 Changes to the Classification Methodology.

We did not include the Price Comparison Prevention DP subclass because it cannot be identified from recordings. We extended the methodology for determining whether a pattern is a dark pattern. As a secondary confirmation measure, Di Geronimo et al. [9] handled disagreements in a pair about DP designations by having a third party cast the final vote; they did this for 40 apps. We applied this measure to all 200 apps in our corpus as the main mode of DP verification. We also aimed to increase the objectivity of this approach by using three different pairs of researchers rather than one pair of two researchers. Time constraints and COVID-19 restrictions forced us to work in pairs online (via Zoom).

## 4 Results

A variety of DPs of varying frequencies across apps and app categories were identified. We provide our anonymous ratings data online[6]. We start by covering the new DP and its subclasses before going into the results in detail.

### 4.1 A NEW DARK PATTERN DISCOVERED IN THE JAPANESE APP CONTEXT (RQ1)

Over the course of analyzing the Japanese apps for DPs, we found a new class of DP: *Linguistic Dead-End.* This is a design pattern wherein language use prevents or makes it very difficult for the user to understand crucial functionality, typically by using unfamiliar words of a foreign origin, even if the local syllabary is used. Importantly, the purveyors of the app would benefit from users not accessing, modifying, or leaving as default these features, functions, and/or information in the app. The presence of proper Japanese elsewhere in the apps suggests that the choices underlying these DPs may have been on purpose, i.e., used in a deceptive way. Moreover, there are many free or affordable translation services available. For example, iubenda[7] and gengo[8] can be used to translate even large and complicated content, such as a TOS, faithfully and with relative ease. Considering these points coupled with a clear benefit to the purveyors, we conclude that these designs were chosen on purpose. In other words, they are dark patterns. We found two varieties, or subclasses, of Linguistic Dead-Ends in the Japanese corpus: *Untranslation* and *Alphabet Soup*. We report on these DPs in detail next. While we were not able to obtain ethics approval to share specific examples in published reports, we have recreated the most common forms of Linguistic Dead-Ends in Figures 1, 2, and 3.

---

[6] Due to ethics restrictions, we can only provide the anonymized version of our data set: https://docs.google.com/spreadsheets/d/1KDqT8EAqjk48MpH65ZNp_kIwdMeC2hFNYezkSJlGpN4/edit?usp=sharing

*Untranslation* refers to a design pattern in which part or all of the app is in a language unfamiliar to the people using it, even if the app is stated as available in the local language in the store (Figure 2). Untranslation DPs prevent or make it exceedingly difficult for the user to access or understand the necessary and/or sensitive information that, by their ignorance, may benefit the purveyor. Notably the apps do not provide a translation function for these sections. A frequent example in our Japanese corpus was the Terms of Service (TOS) being in English. In other cases, almost all functions were in a foreign language/s, even though the app description indicated that the app should be in Japanese. In still others, chat messages sent by the operator were in a foreign language.

The other variety of Linguistic Dead-End, *Alphabet Soup,* covers cases where a word of foreign origin is expressed using the local characters, symbols, or numbers (the correct syllabary), such that the meaning cannot be understood even though the characters can be read (Figure 3). This gives the impression of legitimacy, since the correct syllabary is used. This DP may also exist alongside correct uses of the language. In our corpus, we found cases where certain functions that may be understood in English were written in Japanese katakana, which is often used to represent foreign languages, even though there are legitimate and common Japanese words for these functions. Another example related to money formatting; for instance, ￥1.25 uses the Japanese yen prefix but the use of a period suggests that it is another currency, likely USD. We also found cases where this DP was paired with other DPs, notably Trick Questions and Double Negatives, making the UI even more difficult to understand, even for Japanese users who know some English.

---

[7] https://www.iubenda.com
[8] https://gengo.com

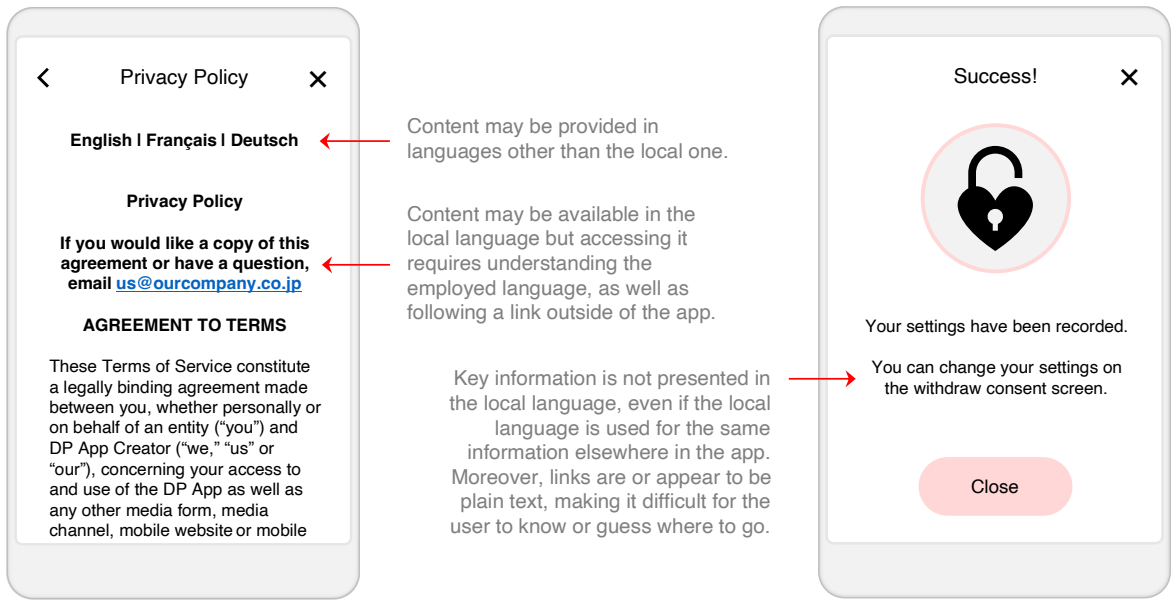

**Figure 2:** *Examples of Untranslation, the most common subclass of Linguistic Dead-Ends in the Japanese app corpus.*

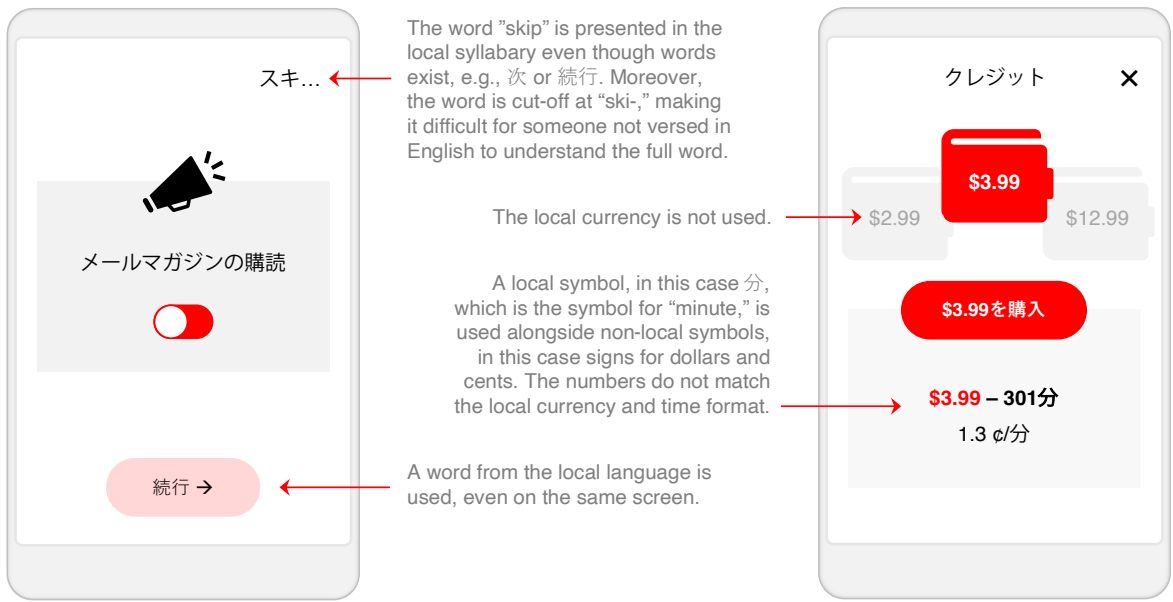

**Figure 3:** *Examples of the Alphabet Soup DP, another form of Linguistic Dead-Ends in the Japanese app corpus.*

Overall, we found 33 instances of Linguistic Dead-Ends across 200 apps, with 29 apps containing at least one Untranslation DP and four with at least one Alphabet Soup DP; of these, two apps contained both. Notably, the distribution of the Untranslation and Alphabet Soups subclasses differed. The app categories with the most Untranslations were, in descending order, 8 in "Communication", 6 in "Photography", 5 in "Music and Audio", 4 in "Family" and "Entertainment", and 2 in "Social." The app categories in which the most Alphabet Soups were found were, in descending order, 2 in "Communication" and 1 in each of "Photography" and "Music and Audio." A paired t-test found a significant difference in the number of Untranslation and Alphabet Soup DPs across app categories, $t(7) = -3.96$, $p = .006$, indicating there were far more cases of Untranslation in our corpus. The lack of coappearance of these subclasses in a single app suggests that they were not deployed as part of a unified DP strategy. Rather, each uniquely takes advantage of deception through linguistic means.

The app developer's country of origin might also have played a role. For Untranslation, countries of origin were: 11 from the USA, five from Korea, four from China, three from Japan, and one from each of Israel, France, Sweden, India, Belarus, and Norway. For Alphabet Soup, two were made in the USA and one each were from China and Israel. Notably, most instances of Untranslation were found in apps created by American developers or Asian nations, including but not limited to Japan. Also, all instances of Alphabet Soup were created by non-Japanese developers. Additionally, the two apps in which both subclasses of Linguistic Dead-End were present were created by a linguistically close Asian nation (China) and a satellite location of a Japanese company (Israel). The numbers are few, so we should take caution when interpreting these results. Yet, they suggest a link between the linguistic savvy of outside nations and deployment of more sophisticated Linguistic Dead-End strategies involving Alphabet Soup or a combination of both subclasses.

We include these new DPs in our taxonomy and analysis, to which we now turn.

### 4.2   DISTRIBUTION BY DARK PATTERN SUB/CLASS (RQ2A)

We first considered the relative distribution of DPs by class and subclass of DP. We begin by describing the cases in which DPs were embedded within the app relative to other UI elements and functions. Table 1 illustrates the DPs cases that we found, what sub/class/es of DP applied to each case, and a qualitative measure of the frequency of occurrence: single (S) or multiple (M). For this, we used the same categories defined by Di Geronimo et al. [9]. By "single" we mean that within one app, there was only one instance of the given DP case. By "one" we mean that either (i) the case occurred only once, e.g., a single popup, or (ii) its occurrence was context-dependent, only found in one place within the app, e.g., always on the checkout screen. By "multiple" we mean that there were multiple instances of the given DP case within a single app, either in terms of (i) frequency of appearance or

(ii) where it could be found, i.e., the context. Note that both S- and M-type DP cases could be found across several apps in our corpus.

**Table 1:** *Specific DP cases, the DP classes or subclasses to which they belong, and their relative frequency.*

| DP Case | Sub/Classes | S/M |
|---|---|---|
| A pop-up appears and interrupts the user in their task | Nagging | M |
| Pop-up to rate | Nagging | S |
| Display of cookies, etc. that appear all the time on some parts of the screen, with no option to turn off the display. | Nagging | S |
| Forced to watch a video for a few seconds when starting the app or after making certain selections | Nagging, Forced Action | S |
| Multiple currencies | Inter. Currency | S |
| When purchasing an item, user will be prompted to purchase more Intermediate Currency than required. | Inter. Currency | S |
| Charging an amount that includes fees not previously shown on the screen just before order confirmation | Sneak into basket | S |
| User clicks a feature (which does not look like a premium) and get a PRO ad or open Apple Store | Bait & Switch, Disguised Ads | M |
| User can't unsubscribe from the newsletter on the app | Roach Motel | S |
| Unable to logout or delete account within the app | Roach Motel | S |
| Terms of service is small and\or grayed out | Hidden Info. | S |
| Terms of use is not displayed during tutorial or member registration | Hidden Info. | S |
| The notifications (and\or emails and SNS) are preselected | Preselection | M |
| The option is preselected | Preselection | M |
| Countdown offer | Toy. with Emot. | M |
| Persuade the user to make the desired choice just before begging for permission. | Toying with Emotions | S |
| There are two or more options, but that is more beneficial for them is more prominent in terms of font, button size, position and frame | False Hierarchy | M |
| Ad appears as normal content | Disguised Ads | M |
| Forced to watch a video for a few seconds when starting the app or after making certain selections | Forced Action | M |
| In the negative form expression of the setting, the check box is not selected, and the one that benefits the company is pre-selected | Trick Questions | S |
| Private settings related dark patterns | Privacy Zuckering | S |
| Words and content in the UI and content are presented in a foreign language, and no translation function is provided in the app. | Untranslation | S |
| The local syllabary is used to express the unfamiliar foreign words even when a word in the local language exists and the rest of the app is in the local language. | Alphabet Soup | S |

Among the 200 Japanese apps in our corpus, 93.5% had one or more DPs, which was close to the 95% found in the US corpus in Di Geronimo et al. [9] An average of 3.9 DPs per app (SD: 2.7, MD: 3) were found compared to an average of 7.4 in Di Geronimo et al. [9]. Figure 1 shows the number of DPs found across all apps. 36.5% of the apps had two or fewer DPs, 46.5% had about 3-6, and 17% had 7 or more DPs.

Each app contained an average of 2.4 (SD: 1.1, MD: 2 ) DPs categorized under one of the six DP classes, including the five classes devised by Gray et al. [16] and the Linguistic Dead-End DPs found in our study. Considering the 16 subclasses outlined by Gray et al. [16], as well as Untranslation and Alphabet Soup, each app contained an average of 2.85 DP subclasses (SD: 1.6, MD: 3). The subclasses of DPs that were found to be more prevalent throughout were, in descending order of frequency, False Hierarchy (n=204), Preselection (n=139), and Nagging (n=138). The bar graph in Figure 1 indicates how much of each subclass of each DP class appeared in each of the 200 apps in our corpus. Most apps (54.5%) contained Preselection, where an option was pre-checked to set a subscription to email newsletters, or where notification permissions were selected before user direction in the notification settings. 52.5% had False Hierarchy, in which the designers intentionally made some choices stand out over multiple other choices. Nagging, found in 43% of the apps, was the third most common DP subclass. This DP includes designs that display ads immediately after app launch or during screen transitions, as well as designs that require app ratings for continued use.

## 4.3   DARK PATTERNS AND APP CATEGORIES (RQ2B)

Next, we considered the relative presence of DPs by app category (Figure 4). Certain DPs can be more prevalent in certain kinds of apps, according to previous research [9]. For example, the presence or absence of a billing component could influence the type and degree of DPs present, and some categories may feature a greater number of paid features or in-app purchase than others, depending on the service model. A Shapiro-Wilk test confirmed that the distribution did not significantly differ from normal ($W$ = .99, $p$ = .63). We then performed an ANOVA test. We found no statistically significant differences between any of the app categories with respect to the distribution of the number of DPs included in the app, $F(7, 192)$ = .88, $p$ = .53. In short, all kinds of DPs could be found across any app category.

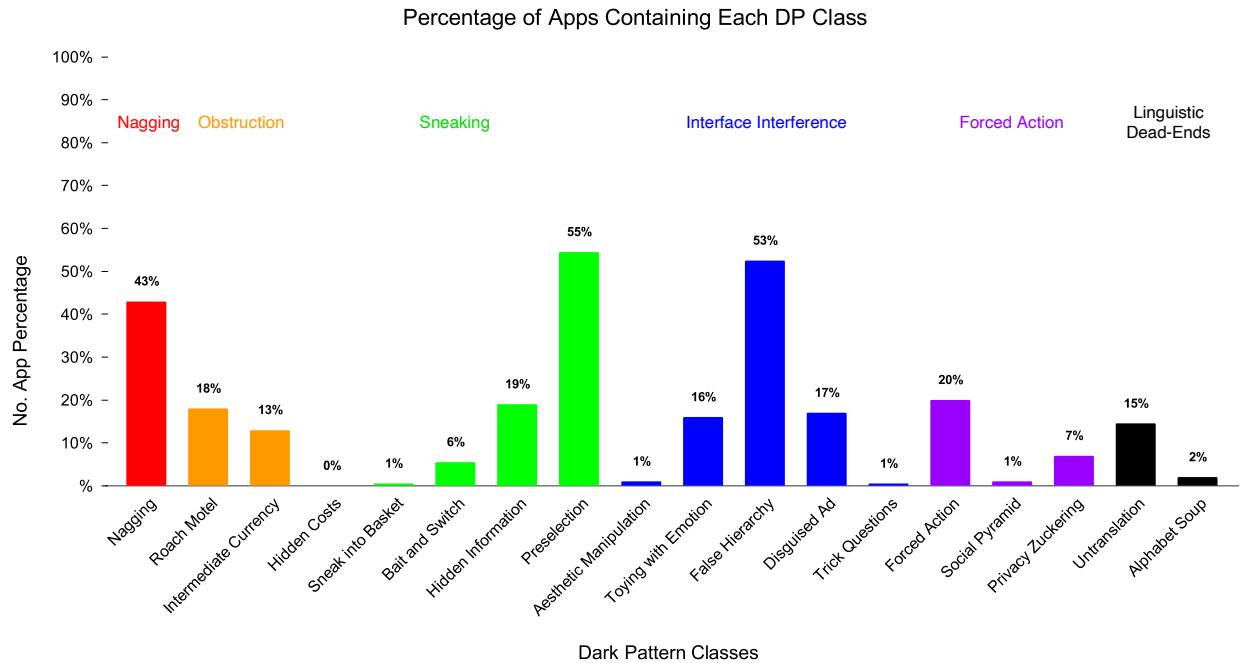

**Figure 4:** *Distribution of DPs among the app categories.*

## 4.4 DARK PATTERNS IN THE JAPANESE VERSUS US CONTEXTS (RQ3)

Since we used the Japanese equivalents of the app store categories covered in Di Geronimo et al. [9], we were also able to compare the relative frequencies of DPs by app category in our corpus and the corpus of Di Geronimo et al [9]. Figure 5 summarizes the results. We note that this is not a controlled comparison, especially given that we used Apple devices and the Apple Store rather than Google-compatible devices and the Google Store. However, as described above, we strived to keep our study setup and corpus of apps as similar as possible, with few deviations outside of the main point of comparison: the cultural context. Indeed, we have little reason to think that the Google and Apple offerings would deviate significantly. Indeed, developers may choose to publish an app on one or both platforms, while other factors, such as revenue models, may play more of a role in UI design choices than the particular type (or category) of app [33].

With these points in mind, we turn to our results on the relative distributions of DPs by context. The values were obtained by calculating the rate of occurrence of each DP in our study minus the rate of occurrence of those DPs given in Di Geronimo et al. [9]. A positive value indicates a higher rate of occurrence in our study, while a negative value indicates a higher rate of occurrence in their study. As the figure indicates, the difference in the occurrence rate of the Aesthetic Manipulation, Privacy Zuckering, and Roach Motel DPs, in that order, was large. On the other hand, Sneak into Basket, Aesthetic Manipulation and Trick Questions, Privacy Zuckering, and Social Pyramid were found in only 1% of the apps in this study, and Hidden Costs was not found in any app.

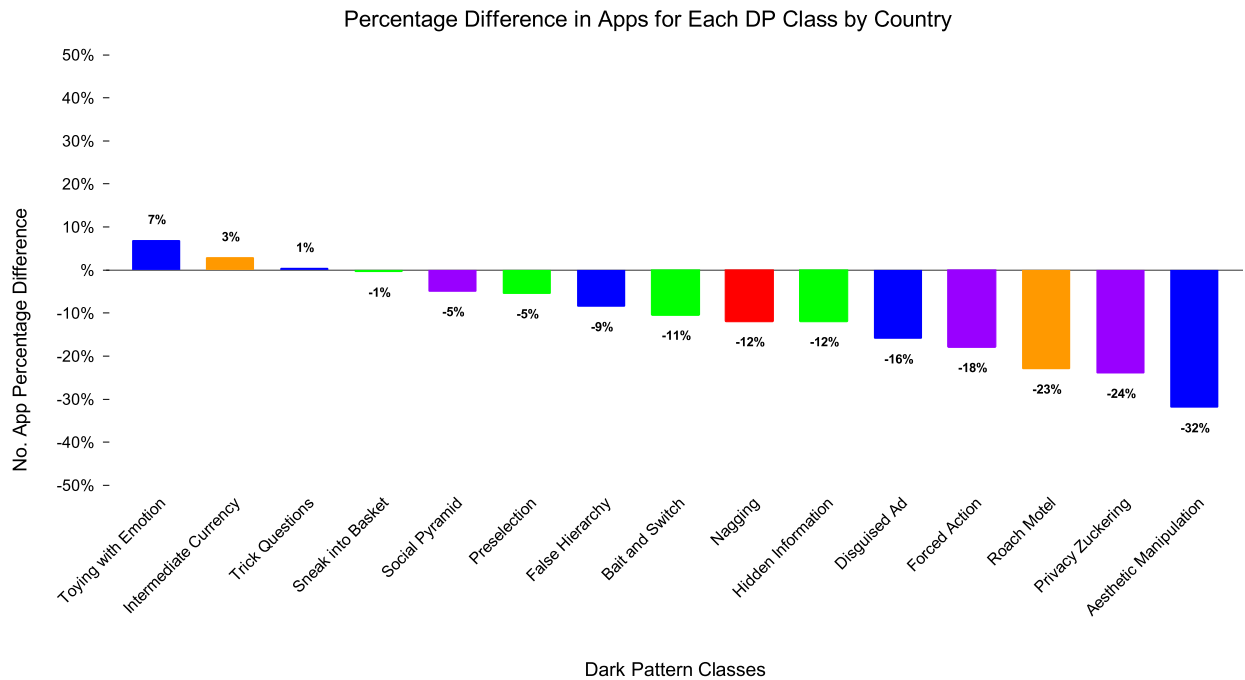

**Figure 5:** *Differences in the distribution of the DPs in our Japanese app corpus compared with Di Geronimo's [9] study.*

## 5 Discussion

In the process of categorizing DPs in apps within the Japanese market, we found that a great number contained malicious patterns that were designed to induce the user to act or not act in specific ways. Additionally, we identified a new class of DP that previous work, largely based in or on the West, was unable to find. Our comparisons of the Western and Eastern contexts also revealed that the occurrence rates of some DPs designs were very different. We now discuss the implications of these results with respect to previous and future work.

### 5.1 A NEW CLASS OF DARK PATTERN: "LINGUISTIC DEAD-ENDS"

A major contribution of this work is the discovery of a new class of DP: the "Linguistic Dead-end," and its subclasses "Untranslation" and "Alphabet Soup." One reason why this DP may have been missed previously is that most DP work has primarily been studied in countries described as WEIRD [19, 20]. Henrich, Heine, and Norenzayan originally found, circa 2010, that 86% of the data used in psychological research samples were from only 12% of the world's population. Although over a decade has passed since this landmark paper, this pattern has recently been found in CHI research. Linxen et al. [24] found that across CHI human subjects studies published between 2016 and 2020, 73% of the participants were from WEIRD countries. Moreover, 52% of all 195 countries in the world had no participant representation in those five years. Yet, as WEIRD research has found, we cannot always generalize the experiences and effects of phenomena across populations, even if we tend to treat WEIRD research results that way. As we described above, in DP research, population samples and national contexts are almost exclusively WEIRD or ambiguous, at best. Given that previous studies of DPs have been biased towards Western contexts, we can say that DP research is not an exception to the general rule. Nevertheless, we need to consider how facets of culture may affect what we know about DPs. Our work is not unprecedented, even within HCI. For example, we know from previous research in a similar context, namely web design, that design aesthetics can be characterized differently depending on culture or country [7]. Our study therefore highlights the importance of cross-cultural studies and the limits of relying on research produced within and for certain nations and cultural contexts. Japan is likely not an outlier; other non-Western and non-English-speaking nations may also be at risk of DPs such as these. This can be confirmed by conducting research in other non-WEIRD countries.

The prevalence of Linguistic Dead-Ends in the Japanese app context also deserves interrogation. One reason could be that foreign-made apps often become a trend in Japan. Considering the apps in which neither Untranslation nor Alphabet Soup were found was also illuminating. For instance, apps developed by Japanese companies dominated the News & Magazines category. Also, in the Shopping category, many fashion apps from Japan that focus on Japanese brands and target, perhaps exclusively, Japanese clientele, i.e., by requiring a Japanese postal address, alongside commercial apps by Japanese companies were at the top of the list. Conversely, the Communication category—the most common category for both Untranslation and Alphabet Soup—included browsers and online communication platforms from competitors all over the world, albeit with English-speaking companies accounting for most of the top spots.

Alternatively, Linguistic Dead-Ends may only exist in certain contexts where the app creators and users have, or are expected to have, some knowledge of different languages. In Japan, for instance, English has long been a feature of public education, and learning English became mandatory for elementary school students in 2020 [13]. App designers may also be taking advantage of "English as the global standard" even while knowing that the intended users may not know English, thus providing most of the app in the language of the users. Yet, this does not explain the overreliance on Untranslation over Alphabet Soup. Perhaps blocks of text are simply easier to manipulate, i.e., translate, especially with machine translation, which improves in naturalness as time goes on. Notably, Alphabet Soup requires a more sophisticated understanding of the local language. Indeed, the pattern we found of outside nations with close cultural and/or linguistic ties, e.g., Korea and China, or inside knowledge of the local tongue, e.g., the satellite office of a Japanese company, deploying both subclasses in one app raises this as a stark possibility.

We must also consider the intersection of legal standards and social expectations. Untranslation, the more prevalent of the two, often (79% of the time) appeared in relation to the app's ToS. A ToS is a contract between the user and the creator/s and stakeholder/s of the app, i.e., a legal construct. Yet, ToS are well-known to be burdensome and confuse laypeople by relying on legalese, jargon, and vague terms, even though Japanese law regulates how they are written. Specifically, Article 3 of Japan's Consumer Contract Law states that there is a need to ensure that the rights and obligations of consumers are clear and plain for consumers[9]. Even so, a study in the U.S. found that 99.6% of ToS are often as long as academic papers [2]. Similarly, in Japan, a study on factors that make ToS difficult to read showed that 36% of 400 respondents agreed to the ToS without reading it, with 123 citing length as the reason [39]. A "darker" reason could be what language the terms are written in; in other words, Untranslation may make the information in the ToS more opaque, further reducing the probability of it being read. Future work, ideally with the creators and purveyors of the Japanese and other non-WEIRD apps that house these Linguistic Dead-Ends will be needed to shed light on the reasons behind their use and the current reliance on Untranslations over Alphabet Soup.

## 5.2 UNDERSTANDING THE DISTRIBUTION OF DARK PATTERNS ACROSS CLASSES AND APP CATEGORIES

A fairly even variety of DP classes and subclasses were found across app categories. A critical look at how the categories may link to the DPs we found may be useful for making sense of these results. The most frequently found DP was Preselection. Notably, the apps that most frequently included this DP did not involve user-generated content, including those found in the Shopping, Entertainment, Social, and News & Magazines app categories. In contrast, apps without this DP either required or encouraged user-created content, such as Photography and Family apps. In other words, in the former category, users consume material that the creators of the app provide, while in the latter category, users themselves use the app to create something. In the former case, it is possible to attract user interest by presenting content through newsletters and notifications. These are contexts in which Preselection can be easily applied. However, in the case of apps that rely on user-generated content, engagement relies on new content that the users themselves create. Defaults and Preselection DPs for encouraging or forcing user-generated content are difficult to imagine for this. These characteristics may have led to the difference in the DP distribution among app categories.

---

[9] https://www.shugiin.go.jp/internet/itdb_housei.nsf/html/housei/h147061.htm

False Hierarchy emerged as the second most commonly found DP. This DP often took the form of a combination of buttons that had prominent colors, such as blue, red, or pink, and buttons that were less prominent, such as those with no outline or with grayed out text. Buttons colored blue are thought to evoke a sense of calm and are often used by companies that have trust as a core value [27], while red is thought to evoke feelings of excitement [23]. Moreover, in the absence of white, blue, red, and vivid colors in general tend to be preferred by Japanese populations [34], and these preferences seem to map onto how False Hierarchy was designed into the Japanese apps in our corpus. We may thus not be surprised that these colors were used in creating visual hierarchies, even within the Japanese context. Given the relative importance of color as a core element of visual UIs, we may also not be surprised to find so many instances of a DP that relies on how color is perceived by the user.

The third most common DP was Nagging, which is a very common DP the world over. As Mathur, Kshirsagar, and Meyer [29] explain, Nagging imposes a cognitive burden on the user, with the goal of wearing them down until they accept whatever option alleviates the burden. In some sense, Nagging may be considered a neutral DP, applicable to any app regardless of category or purpose, because its value lies in pressuring the user to engage with the app in any way that the creators define. In other words, this DP is not tied to a specific goal, task, activity, context, theme, technical requirement, and so on. Given the relatively fewer frequency of DPs in general compared to Western rates, it may be easier for app creators working in the Japanese context to create and widely deploy such generally applicable DPs.

## 5.3   BUILDING ON THE FINDINGS OF DI GERONIMO ET AL. [9] CROSS-CULTURALLY

The work of Di Geronimo et al. [9] formed the basis of this research. Indeed, we attempted to replicate their protocol to discover how DPs operate within the Japanese context and outside of the Google suite of devices and apps. Unexpectedly, we found far fewer DPs overall. Differing app store guidelines and laws could help explain this. We were also inspired to undertake this research because DPs seem not to be widely known or discussed in Japanese circles yet; this could also indicate that app creators in the Japanese market are only just starting to recognize and employ DPs. In other words, DPs may be novel in Japan, and only time will show if this difference is a feature of that novelty or not.

At present, can focus on the known issue of app store choice [33], which was also one of the main differences between our study and that of Di Geronimo et al. [9]. We found that Aesthetic Manipulation, Privacy Zuckering, and Roach Motel, in that order, showed the largest difference in occurrence rates. Aesthetic Manipulation, which was rare in our Japanese corpus, was found in 33% of the apps in the US corpus [9]. This may be due to differences between Google and Apple app store publication guidelines. According to Apple's App Store Review Guidelines:

> *Interstitial ads or ads that interrupt or block the user experience must clearly indicate that they are an ad, must not manipulate or trick users into tapping into them, and must provide easily accessible and visible close/skip buttons large enough for people to easily dismiss the ad. [1]*

In other words, use of Aesthetic Manipulation DPs, such as moving ads and the use of small buttons for closing ads,  will not pass the App Store review process, even if they may be allowed by Google.

Privacy Zuckering was found in 31% of the apps used in the previous study [9], while it was found in 7% of the apps in this study. For apps found to have Privacy Zuckering in this study, the ToS required permission to obtain user information. In other words, the app was designed in such a way that the provision of user information was essential to use the app, and thus it was recognized as a DP. Other apps that attempted to acquire users' app usage data had their permission requested by Apple's uniform pop-up. The reason why this study found a very small number of this dark pattern compared to the study by Di Geronimo et al. [9] may be Apple's criteria for app applications:

> *Apps that collect user or usage data must secure user consent for the collection, even if such data is considered to be anonymous at the time of or immediately following collection. [1]*

## 5.4 IMPLICATIONS

Our findings confirm and extend the scholarship on DPs in key ways that may have implications for researchers, practitioners, and the general app-using public.

### 5.4.1 Cross-cultural research on DPs is needed.

As the first known study on DPs in Japan, this work represents a first step towards a global understanding of DPs. We drew from existing research, using a qualitative expert evaluation protocol to explore the distribution of DPs for the case of Japan. This allowed us to make comparisons with results from other countries. We found that DPs exist in the Japanese app context, and in no small numbers or variety. User perceptions of and reactions to design in general and DPs specifically may differ between the West and the East, broadly speaking, or by cultural context, more pointedly. There is much room for verification in Japan and other WEIRD and non-WEIRD nations for the greater goal of knowledge production as well as the ethical goal of reducing any harms, potential or actual, caused by DPs or arising from their presence with the larger app context. In addition, considering the ethical murkiness of DPs, it will be important to collaborate with professionals and companies that work on UX with integrity.

### 5.4.2 Cross-cultural research on DPs should address language broadly.

Mobile apps may be used by a variety of people who speak different languages and speak a given language to different degrees across a range of countries. When translating or localizing apps, it is necessary to consider possible cultural differences from a linguistic angle: sentences, vocabulary, syllables, numbers, currency, and so on. User-centred design has a role to play here. For example, Japanese and Korean have the concept of "honorifics," which are word endings, prefixes, or phrases that change depending on who you are speaking to. These can indicate social hierarchies, respect or a lack thereof, formality or informality, and even psychological closeness [38]. If an app developed in a country without the concept of honorifics is used in a country with the concept of honorifics, the translation will not take this into account and could result in unintentionally influential designs, such as strong tone of voice anchored to particular user prompts, or even DPs, such as Toying with Emotion. As the new DP and its subclasses reveal, language can be used to persuade and manipulate end-users. Japanese (and Korean) are just the beginning: future work will need to explore other languages and syllabaries for the same or new DPs embedded or reliant on how interactions are linguistically expressed.

### 5.4.3   *Cross-cultural research on DPs should go beyond language.*

While our main results related to replicating patterns found outside of the Japanese context and patterns related to the Japanese language, we also found indications that other features of cultural expressed in design and user interactions have a role to play. A key example is colour and how it was used to create False Hierarchies: the same DP, but with a Japanese flavour. We expect there to be similar "flavours" for other DPs that may only be found through direct comparison of UIs and interactions, which we could not do in this work. Future work may take on such direction comparisons, especially of "the same" app, e.g., Uber, which is available in several countries and languages. Future work can also involve controlled user studies involving a diversity of participants, including consumers and design experts, who may be able to identify lower-level differences in how DPs are visually or interactively expressed. This would add a new, cross-cultural dimension to the work already being done with user studies of DPs [3, 4, 9, 14].

## 5.5   LIMITATIONS AND FUTURE WORK

This work has several limitations. The new DPs that we found were only assessed within the Japanese context; future work will need to examine other language contexts to determine whether and to what extent these DPs are found elsewhere. We also only considered apps for smartphones and mobile devices; future work will need to consider other technology setups, including websites of various kinds, software, and so on. Although we had planned to, were unable to obtain ethics approval to disclose our list of apps or conduct human subjects research, due to concerns over published criticism of a large corpus of commercial and non-commercial property. Future work should seek to understand the effects on real or potential end-users through participant-based research methods. Future work can also involve working with companies and designers directly, carrying out interviews with the people creating and deploying DPs about their ethical stances, whether they agree theses design patterns are DPs, and what benefits they receive or expect to from their deployment. Academic-industrial collaborations could also work together to investigate the nature of DPs on end-users and consumers through nonbranded but ecologically valid versions of DP-containing apps as research instruments. Finally, developers and companies could provide a list or means of deriving and accessing a record of app versions so as to chart out a history and timeline of DP integration for a single app and how each DP within the app changed over time. This could provide fine-grained contextual information of the spread of DPs and trends in DPs, which could be compared cross-culturally. Rather than take on this laborious and time-consuming task manually, industrial collaboration or other means, such as automated detection or crowdsourcing, could enable this line of research.

## 6 Conclusion

Dark patterns have become a hot topic in critical computing and interaction design at large. In this work, we have provided evidence that they exist at a broader scale in the case study of Japanese apps. We have also shown the value in approaching research on DPs from a cross-cultural perspective, identifying a new form of DP that may only be found with a culturally- and linguistically-sensitive lens. Future work will need to explore whether and how this DP exists in other contexts beyond the West and in languages other than English. Moreover, the Japanese-specific properties of this DP and its subclasses suggest the possibility of other, still hidden DPs within other cultural and linguistic contexts that may only be unearthed by targeted, critical work. Critical computing research, ethical design praxis, and the global e-commerce sector may benefit from cross-cultural and intersectional initiatives within and beyond apps.

### ACKNOWLEDGMENTS

We thank the members and staff in the Aspirational Computing Lab for supporting this work and providing early feedback. This work was funded by departmental funds.